\documentclass[letterpaper,twocolumn,10pt]{article}
\usepackage{usenix}

\usepackage{tikz}
\usepackage{amsmath}
\usepackage{amsthm, amssymb}
\theoremstyle{definition}
\newtheorem{definition}{Definition}[section]

\usepackage{multirow}
\usepackage{filecontents}
\usepackage{graphicx}
\usepackage{subfigure}
\usepackage{epstopdf}
\usepackage{booktabs}
\usepackage{algorithmicx}
\usepackage[Algorithm,ruled]{algorithm}

\usepackage{soul}

\usepackage{xspace}
\newcommand{\name}{\texttt{SecCode}\xspace}

\newcommand*\circledwhite[1]{%
  \tikz[baseline=(char.base)]{
    \node[shape=circle, draw, fill=black, text=white, inner sep=-0.1pt] (char) {#1};}}

\usepackage{balance}
\definecolor{codegreen}{rgb}{0,0.6,0}
\definecolor{codegray}{rgb}{0.5,0.5,0.5}
\definecolor{codepurple}{rgb}{0.58,0,0.82}
\definecolor{backcolour}{rgb}{0.95,0.95,0.92}

\usepackage{listings}
\usepackage{xcolor}
 \lstdefinestyle{mystyle}{
  backgroundcolor=\color{backcolour}, commentstyle=\color{codegreen},
  keywordstyle=\color{magenta},
  numberstyle=\tiny\color{codegray},
  stringstyle=\color{codepurple},
  basicstyle=\ttfamily\footnotesize,
  breakatwhitespace=false,         
  breaklines=true,                 
  captionpos=b,                    
  keepspaces=true,                 
  numbers=left,                    
  numbersep=5pt,                  
  showspaces=false,                
  showstringspaces=false,
  showtabs=false,                  
  tabsize=2
}

\usepackage[utf8]{inputenc} 
\usepackage{tcolorbox} 

\usepackage{subcaption}

\usepackage{listings}
\usepackage{xcolor}
\usepackage{babel,blindtext}

\usepackage{authblk}
\makeatletter
\renewcommand\AB@affilsepx{, \protect\Affilfont}
\makeatother
\begin{document}
%-------------------------------------------------------------------------------

%don't want date printed
\date{}

\title{From Solitary Directives to Interactive Encouragement! \\ LLM Secure Code Generation by Natural Language Prompting}

%\author{
%{\rm Anonymous}
%}

%\affil[1]{\small \rm CSIRO'Data61} 
%\affil[2]{\small \rm Swinburne University of Technology}
%\affil[3]{\small \rm DST Group, Australia}

\author[1,*]{\rm Shigang Liu}
\author[1,*]{\rm Bushra Sabir}
\author[1]{\rm Seung Ick Jang}
\author[2]{\rm Yuval Kansal}
\author[1]{\rm Yansong Gao}
\author[1]{\rm Kristen Moore}
\author[1]{\rm Alsharif Abuadbba}
\author[1]{\rm Surya Nepal} 

\affil[1]{\rm CSIRO's Data61, Australia}
\affil[2]{\rm Princeton University, USA} 
\affil[*]{\rm These authors contributed equally to this study.}

\maketitle
\pagestyle{empty}

%-------------------------------------------------------------------------------
\begin{abstract}
%The field of Large Language Models for Code (Code LLMs) is rapidly expanding, demonstrating impressive code generation capabilities. While existing literature predominantly focuses on functional correctness, the security aspects of generated code have received less attention. As the adoption of LLM-generated code increases, understanding its security implications becomes crucial, extending to LLMs' potential role in enhancing code security. Although preliminary evaluations have deemed LLMs unsuitable for vulnerability identification and remediation, their full potential in addressing security issues remains underexplored. Beyond security concerns, the usability of these models also warrants consideration.  Current LLM-based code generation approaches typically employ prompts consisting of either pure programming language (i.e., code) or a combination of natural language (NL) and code. While code-only prompts offer specificity, they may present challenges for users less versed in programming. The efficacy of pure NL prompts, despite their potential accessibility, remains understudied, presenting an opportunity to explore more user-friendly interfaces for LLM-driven code generation.
% The exclusive use of programming language code can be less user-friendly for typical LLM users, and there has been no exploration of the potential of NL prompts, despite their ease of use.
%
Large Language Models (LLMs) have shown remarkable potential in code generation, making them increasingly important in the field. 
However, the security issues of generated code 
%should not be overlooked
%largely been overlooked, with the existing literature predominantly focusing on functional correctness. Additionally, 
have  not been fully addressed, 
and the usability of LLMs in code generation still requires further exploration. Current approaches often use prompts composed of either code or a mix of natural language (NL) and code. Although 
%code-involved 
prompts involving code might offer precision, they can be difficult for users with less programming experience. The potential of using purely NL prompts for secure code generation remains unexplored, highlighting the need for more user-friendly, NL-based interfaces for LLM-driven secure code generation. %\footnote{*{These authors contributed equally to this study.}}

%To address the aforementioned limitations, this work introduces \name, a framework that leverages an innovative interactive encouragement learning technique for secure code generation using solely NL prompts, which can be publicly shared with common users. \name operates through three stages. 1) Code Generation using NL Prompts: \name generates code based on user-friendly NL prompts. 2) Code Vulnerability Detection and Fixing: Utilizing encouragement learning, \name detects and fixes vulnerabilities through the LLM alone. 3) Vulnerability Cross-Check and Code Security Refinement: \name incorporates external feedback from rule-based software vulnerability detection tools into dynamically updated LLM prompts to refine code security. These stages are executed in multiple iterative loops to enhance security, rather than relying on a naive single execution.

To address the aforementioned limitations, this work introduces \name, a framework that leverages an innovative interactive encouragement prompting (EP)  technique for secure code generation with \textit{only NL} prompts. This approach ensures that the prompts can be easily shared and understood by general users. \name functions through three stages:
1) Code Generation using NL Prompts; 2) Code Vulnerability Detection and Fixing, utilising our proposed encouragement prompting; 3) Vulnerability Cross-Checking and Code Security Refinement. These stages are executed in multiple interactive iterations to progressively enhance security. 
% We evaluated \name in different sceMisplaced alignment tab character &.narios to demonstrate its effectiveness\Garrison{we mention LLMs and datasets (or number of datasets) used here.}.
By using both proprietary LLMs (i.e., GPT-3.5 Turbo, GPT-4 and GPT-4o) and open-source LLMs (i.e., Llama 3.1 8B Instruct, DeepSeek Coder V2 Lite Instruct) evaluated on three benchmark datasets, extensive experimental results show that our proposed \name greatly outperforms compared baselines, generating secure code with a high vulnerability correction rate. For example, \name exhibits a high fix success rate of over 76\% after running 5 automated EP interactive iterations and over 89\% after running 10 automated EP interactive iterations. To the best of our knowledge, this work is the first to formulate secure code generation with NL prompts only. We have open-sourced
%\Garrison{Be careful about this statement if we do not put the source code link here. we have this: "anonymous according to the submission policy"} \Shigang{I think that's ok, I used this quite a lot in my previous papers, no reviewer raise any question about this}\sharif{I asked Bushra to prepare anonymous repo. If ready would be good to include in contributions. It makes it stronger :)} 
our code and 
%call for attention to secure code generation within 
encourage the community to focus on secure code generation.

%\textcolor{red}{We can highlight two facts upon experimental results, both for the first time! The SecCode itself is indeed a good vulnerability detector and fixer through interactive encouragement prompt design. The SecCode can generate functional \textit{and} secure code by feeding a NL prompt alone.}

%To show the effectiveness of the proposed SecCode, we evaluated it in different scenarios, including comparisons with a series of baselines, such as prompt-based LLMs and evaluations with different LLMs over various \hl{established benchmark} datasets for secure code generation. Experimental results show that our proposed SecCode exceeds the baselines and is able to generate secure code with a high vulnerability correction rate. For example, our experimental results show that SecCode can achieve a successful correction rate of over 76\% after running 5 automated feedback loops and over 89\% after running 10 automated feedback loops. \hl{To the best of our knowledge, this work is among the first {\color{red} [Kristen - have we  we're not the first?]} to formulate secure code generation with NL prompts only, reflecting the real-world need for accessible and easy-to-use tools for the wider developer community}. We open-sourced our code and call for attention to secure code generation within the community.
%We believe that this work will serve as a stepping stone for future research in secure code generation.
\end{abstract}

\section{Introduction}

% \textcolor{blue}{Remark of discussion between Sharif and Garrison: The structure can firstly point out two limitations of state-of-the-art methods: turning LLM into an effective code vulnerability detector and fixer upon only the prompt while keeping the underlying LLM intact; generating secure code by only NL prompt. Then we resolve this by proposing SecCode. However, we face three challenges. Then we state how we address each of these three challenges by mapping each stage to one of the challenges.}

% \textcolor{blue}{In addition, we can use three words: iteratively to refer to the loop; collaboratively to refer to harnessing human intelligence of designing the prompt, LLM intelligence and external knowledge from the CodeQL; automated refer to that once the prompt is devised, the entire process of SecCode will be automated running.} \Shigang{Yes, I agree. Please feel free to revise the paper. I will continue to work on it the week after.}

%\Garrison{When referring to encouragement learning, use `encouragement prompting' or `encouragement prompt'. The latter two are exchangeable. I have gone through Section III and changed all the terms accordingly, Section 2.1 has also been changed. } 

Software vulnerability analysis has long been a critical aspect of software development \cite{le2022survey}. This challenge has intensified with the widespread adoption of large language models (LLMs) for code generation. These models include both closed-source solutions (e.g., commercial code completion engines) and open-source ones (e.g., academic projects). For example, the Copilot commercial tool is used by more than 1 million developers and more than 5,000 businesses \cite{he2023large}. Similarly, the open source tool CodeGeex boasts tens of thousands of active daily users, each averaging more than 250 API calls per week \cite{zheng2023codegeex}. This broad acceptance and utilization of LLMs for code generation underscores their role in enhancing programming productivity.

\begin{table}[!t]
\centering
\caption{Experimental results based on NL prompt for SecEval and Holmes datasets using GPT-3.5. Vul.\_Gen.: indicates the percentage of vulnerable samples. %\hl{FSR refers to the fixed success rate, as defined in Section} \ref{sec:EM}. 
%for the generated code; FSR means the percentage of fixed vulnerable samples.
}
\begin{tabular}{|l|l|l|c|c|}
\hline
Data    & \#prompt & Prog. Lang. & \multicolumn{1}{l|}{Vul.\_Gen.} & \multicolumn{1}{l|}{Fixed} \\ \hline
SecEval & \ \ \ \  213      & Python      & 59\%                      & 43\%                     \\ \hline
Holmes  & \ \ \ \  121      & CPP/ Python & 64\%                      & 5\%                      \\ \hline
\end{tabular}
\label{Table1}
\end{table}

It is widely acknowledged that most LLMs do not address security issues during code generation but focus merely on improving their code functionality. Generally, code generation LLMs are pretrained models that use prompts as inputs to produce code from various programming languages (e.g., C or Python) as outputs. This setup lacks consideration of security mechanisms, often resulting in vulnerabilities in the generated code. For instance, a recent study found that 40\% of the generated programs contain serious vulnerabilities, including those generated by ChatGPT \cite{he2023large}. 
Table \ref{Table1} presents our pilot experimental results using GPT-3.5 based on our constructed NL prompt datasets\cite{tony2023llmseceval, ullah2024llms}. 
% The second column lists the number of prompts, the third column denotes the programming language, the 
The second last column shows the percentage of vulnerabilities in the generated code, and the last column shows the percentage of vulnerable samples that can be fixed by the basic prompt \emph{``Now you are a security expert, please identify the possible MITRE TOP 25 vulnerabilities in the following code snippet and fix them thoroughly.''}
Our experiments highlight that more than half of the generated code samples in both datasets contain MITRE TOP 25 vulnerabilities, with SecEval and Holmes datasets showing vulnerabilities in 59\% and 64\% of samples, respectively. The results also showed that these insecure codes can not be fixed by using the naive prompt. This underscores the critical importance of addressing vulnerabilities in generated code to ensure software security.

\noindent{\bf Challenges:} While numerous techniques have been developed to identify security vulnerabilities in AI-generated code, research specifically focused on \textit{securing} the code generation process remains limited, 
% \textcolor{red}{does not make sense, you said many, then said few. Do you mean many working on code completion or code based prompt? few working on NL based prompt?} \Shigang{I mean - Many works have investigated the security issues of code generation using AI, but only a few have proposed solutions. In other words, there is a lack of solutions for secure code generation}. 
indicating that secure code generation is still an emerging field. Khoury \textit{et al.} \cite{khoury2023secure} noted that LLMs like ChatGPT can potentially identify code vulnerabilities,
%\Garrison{do you mean the raw code instead of code generated by LLMs?}
but code generated by LLMs is often not robust against certain attacks. 
SVEN \cite{he2023large} is proposed to enhance the capability of LLMs in generating secure code. However, SVEN relies on a mixture of programming language and natural language prompts, and there is significant room for improvement in HumanEval performance in terms of functional correctness. 
%Liu \textit{et al.} \cite{liu2024your} introduced EvalPlus, a framework to evaluate functional correctness and identify erroneous code synthesized by LLMs. 
Very recent studies \cite{ullah2024llms, zhou2024large, liu2024your} 
% indicate that insecure code generated by AI assistants 
% \textcolor{red}{Is AI assistant referring to LLMs or something else?} \Shigang{According to recently published papers, this refers to Artificial Intelligence-related tools, specifically LLMs, as they are one of the most popular techniques for code generation. Honestly, an experienced reviewer will not raise any question about this, we can ignore this part, no need to change anything} 
% are challenging to be detected and fixed also by these AI assistants or LLMs. 
highlight that, in terms of security, AI assistants and LLMs struggle to detect and fix insecure code that they themselves generate.
% and resolution of software vulnerabilities. 
Therefore, there is a pressing need for novel methodologies to secure LLM-powered code generation using only natural language prompts while satisfying the practical demand for user-friendly, accessible tools within the broader developer community. To achieve this, there are three identified major challenges.

\noindent$\blacksquare$\textit{ Challenge \#1: Scarcity of Diverse Natural Language-Only Prompts for Secure Code Generation.} %Programming Language Prompts vs Natural Language Prompts.
% Due to a large number of AI assistant tools \cite{he2023large, zheng2023codegeex, deepseek-coder, nye2021show} available for generating programming language code, software developers typically rely on these tools for code generation. 
Most AI assistants and LLMs \cite{choi2023codeprompt, wei2022chain, chen2022program, nye2021show} generate code using one of two approaches. Either relying on prompts written in a programming language or alternatively using a mixture of programming language and natural language as prompts, based on the developer's request.
However, there is a lack of studies focusing on code generation using natural language prompts as input. In addition, to be user-friendly, the requested user-provided prompt must be \textit{simple}, despite that this seed prompt can be automatically augmented. Moreover, there is a notable scarcity of publicly available datasets featuring prompts designed solely in NL for software code generation. We note that there is only one such dataset of LLMSecEva, which exists~\cite{tony2023llmseceval}, focusing on MITRE's top 25 CWEs released in 2021. The CWEs are lack of diversity, which is prohibitive for comprehensive evaluations.
% This absence poses a significant obstacle to research and development in the community. 
%Consequently, the functionality and security of code generated through this approach remain unexplored. 
% A recent study demonstrated the usefulness of generating source code snippets directly from natural language prompts \cite{xu2022ide}.
Therefore, the first challenge is to figure out \textit{how to utilize simple user-provided seed NL prompts rather than programming language prompts to generate secure software code, compounded by the lack of available benchmark datasets.}
%in order to 
% enabling a better understanding of code functionality and security. 
%This challenge is compounded by the lack of public datasets fitting this evaluation demand.}
%Therefore, the first challenge is to determine \textit{how to design NL prompts to generate software code, in order to understand the code functionality and security?}
% where we could examine the security vulenrabilities
% \textcolor{red}{You mean NL prompts or hybrid of programing and NL prompts?} \sharif{We need to close the paragraph with the challenge that we need to address? We can say, "However, there are no existing designed NL solely prompts for generating software code where we could examine the security vulenrabilities!!​". Therefore, "How can we effectively design realistic prompts to generate software code and be able to examine the top vulnerabilities identified in the MITRE top 25 list?"} \Shigang{please feel free to make changes, I will double check later}
%Therefore, developing an effective method using NL prompt for code generation is a highly challenging problem.

\noindent$\blacksquare$ \textit{Challenge \#2: Ineffective Prompt Strategies Undermine LLMs' Potential for Code Vulnerability Analysis and Remediation.} 
% Generic Vs Precision-Tailored Prompts.
Emerging LLMs \cite{lyu2024automatic}, including Codex, AlphaCode, CodeGen, DeepSeek Coder, GPT-3.5, and GPT-4 are being developed for code generation. However, research has raised concerns about security vulnerabilities in LLM-produced code, including the use of untrustworthy libraries. Additionally, many researchers \cite{ullah2024llms, liu2024your, wu2023effective, becker2023programming} find that there are often security issues in the AI assistant generated code.
Recent studies indicate that LLMs are unreliable in reasoning about software vulnerabilities based on simple prompts, often leading to incorrect and unfaithful conclusions \cite{ullah2024llms}. Another study found errors in code generated by ChatGPT \cite{liu2024your}. Wu et al. \cite{wu2023effective} noted that LLMs themselves are ineffective in detecting software vulnerabilities. 
%\hl{We hypothesize this limitation arises from the use of simple, one-shot prompts that are shallow and directive, querying the LLMs only once and drawing conclusions, without fully leveraging their potential for detecting and fixing code vulnerabilities}. 
Specifically, prompts like chain-of-thought are not optimized for code generation or software vulnerability patching. 
The research highlights the need for further advancements in developing domain-specific prompts for LLMs, ensuring they can effectively and reliably function as code vulnerability detectors or evaluators.
Therefore, \textit{designing effective interactive prompts for accurate software vulnerability analysis, as well as efficient patching of these vulnerabilities, poses a very challenging problem.}

\noindent$\blacksquare$ \textit{Challenge \#3: Inadequate LLM Self-Alignment and the Critical Need for Rigorous Cross-Checking.}  %Guiding Light for LLM - Cross-Assessment}
%\Garrison{We need to use the term consistent, in many places including Abstract, we use `Cross-Check'.}
%\hl{Dilemma Between Self-Alignment Assessment and Cross-Assessment} } \sharif{I feel there are work already using CodeQL, right? so our pitch could be LLMs alone might not be the best. Also static analysis tools have been there for a while and might have false-postive/negagtive, whatever issues? and we make the argument for hybrid approach here} \Shigang{Maybe we shouldn't mention false positives and false negatives because LLM and CodeQL also have these issues.}\sharif{Ok. Sure. Phrase as u think relvent and remove the color text. Thanks!}
% Self-Alignment vs Cross-Alignment
Alrashedy and Aljasser \cite{alrashedy2023can} emphasized that LLMs alone are insufficient for patching software vulnerabilities. However, performance improvements can be achieved when LLMs are guided by feedback for code refinement~\cite{alrashedy2023can}. This suggests that relying solely on LLMs for vulnerability detection and software vulnerability patching is not advisable, as they can  potentially produce misleading and false information~\cite{shen2023large}. For instance, an LLM might incorrectly overlook potential vulnerabilities in order to produce what appears to be a `successful' code vulnerability detection. 
%Aligning the LLM behavior with cross-checking mechanisms or human feedback has been widely accepted to resolve this problem in other domains~\cite{shen2023large}.
As acknowledged by previous studies \cite{ullah2024llms, wu2023effective, alrashedy2023can}, there is often a significant improvement in secure code generation when advanced feedback is provided. In this research context, He and Vechev \cite{he2023large}, who focused on security issues in generated code, leveraged CodeQL---a rule-based software toolkit for code vulnerability static debugging and detection---to perform security checks, achieving reliable results. However, their work involved a mix of programming language and natural language as input, and lacked specifically designed prompts for secure code generation, primarily focusing on functional correctness rather than fix success rate.
%Securing code generation is not a single-step task. It involves conducting a comprehensive security check of the generated code, and if vulnerabilities are detected, feeding this information back to the LLM for further refinement. This generation/correction, security check and information feedback can be run multiple interactive iterations rather than terminating the process after a single iteration. 
Therefore, \textit{designing a novel pipeline that provides alignment and then cross-checking for LLMs in software vulnerability detention and patching for secure code generation is a challenging problem.}

% \Garrison{We have not pointed out the research gap compared to existing works to this end.} \Shigang{We are discussing the challenge here and have not yet mentioned the solution, so this may not be the right place to highlight the gap.}

% In this case, the key challenge is identifying what kind of advanced feedback the model can provide to the code generation LLM that is of vital importance for generating secure code.

\noindent\textbf{Our solution:} To tackle the three challenges, we propose the \name scheme for secure code generation. \name includes three stages to tackle the above three challenges:  
1) Code Generation using simple user-provided NL Prompts: \name generates code based on user-friendly NL prompts. These simple prompts are automatically enhanced by the \name pipeline. To address the scarcity of datasets containing NL prompt and code pairs covering a wide range of CWEs, we have constructed two additional datasets for comprehensive evaluations (\textbf{Challenge \#1)}.
2) Code Vulnerability Detection and Fixing: Using the novel encouragement prompting and interactive feedback with our proposed LLM-Optimized prompts (\textbf{Challenge \#2)}. This method moves beyond a one-shot, directive approach, by developing the first method for encouragement prompting and interactive feedback. Our approach engages LLMs in a series of interactive iterations, designed to fully leverage their potential for vulnerability detection and remediation.
% which stimulates LLMs over a series of interactive iterations to leverage their maximum potential for vulnerability detection and fixing. 
Specficially, we incentivize LLMs to generate secure code by providing vulnerable program code and the relevant vulnerability information. We implement a prompt-based encouragement strategy and propose an interactive iteration strategy to automatically check and fix security issues in the generated code using LLMs. 
3) Vulnerability Cross-Check and Code Security Refinement: \name incorporates external feedback from rule-based software vulnerability detection tools, feeding it back into the LLM through dynamically updated prompts to continuously refine code security (\textbf{Challenge \#3)}. That is, we implement a new strategy that integrates the LLM self-alignment with CodeQL for advanced feedback generation. This process cross-checks any possible existing vulnerabilities and then provides additional information for the LLMs to further correct the security issues in the generated code.
In other words, securing code generation is not a single-step task. It involves conducting a comprehensive security check of the generated code, and if vulnerabilities are detected, feeding this information back to the LLM for further refinement. This generation/correction, security check, and feedback process can be executed over multiple interactive iterations rather than stopping after a single iteration.
It is worth noting that we are the \textbf{first to propose encouragement prompting} for secure code generation, and we are among the \textbf{first to propose the use of the interactive iteration prompting and the combination of LLMs with CodeQL} to augment
% \hl{trigger} \Shigang{or we could replace trigger with: use CodeQL as a guiding light to LLM for secure code refinement}
LLMs for secure code refinement. Our experimental results show that \name outperforms the baselines in terms of vulnerability correction rate, as well as functional correctness.
\noindent{\bf Contribution.} Our main contributions are as follows.

\noindent$\bullet$ We are the first to propose the novel Encouragement Prompting  (EP) for secure code generation. Our experiments show that EP can help LLMs achieve significantly better performance in secure code generation. 

\noindent$\bullet$ We are the first to propose the strategy of interactive iteration prompting in providing the vulnerability information and fixing the vulnerability automatically. 

\noindent$\bullet$ We propose a hybrid refinement strategy that integrates both LLMs and CodeQL to provide comprehensive feedback and cross-checking. This approach incorporates CodeQL into the gamification iteration, delivering enriched feedback to the LLMs for enhanced secure code refinement. 
%{\color{red} Maybe we can introduce "gamification" earlier if we want to use that expression? Eg. in "Our Solution" part 2.}

\noindent$\bullet$ We develop and implement the \name framework. We have conducted extensive experiments using a range of proprietary LLMs in different scenarios.
We have open-sourced the code\footnote{\emph{Will be available soon}} and prompts for other researchers to evaluate and further contribute to this research area.

\section{Preliminaries and Related Work}

\subsection{Preliminaries}

This section introduces the preliminaries: the prompt engineering definitions and the secure code generation. \\

%\subsubsection{Prompt Engineering}
\noindent\textbf{Prompt Engineering.} 
Prompt engineering is crucial in software security, especially for secure code generation, as it ensures precision and clarity, making the generated code meet specific security requirements. By including relevant context and explicitly requesting security practices, prompt engineering helps the model understand and adhere to security constraints. Most existing work primarily focuses on code generation, while secure code generation using purely NL prompts remains in its early stages in both academia and industry. As far as we know, we are the first to propose a pipeline for secure code generation using purely NL prompts.
%this study endeavours to fill the gap of code generation and . 
To achieve this, we propose encouragement prompting and adaptive prompting
%{\color{red} [Do we want to put 'adaptive learning' into the intro as well?]} \Shigang{probably not this is optional for the users, not part of our pipeline} 
for enhancing secure code generation. 

\begin{definition}
  \textbf{Encouragement Prompting (EP)}: Encouragement prompting is an award-driven approach that enhances a model's capabilities by providing motivational feedback and incentives.
\end{definition}

In EP, we use a scoring system as a form of reward. For example, the model receives \textbf{1 point for each vulnerability fixed, and 1 point as a bonus if all vulnerabilities are successfully fixed.} The following is an example structure regarding EP in this work:

\begin{tcolorbox}
\textbf{Example of EP:} \\
You are a software security expert, your goal is ... \\
Goal Instructions: \\
(a) | (b) | (c) | (d) | (e) | etc. \\
 YOUR REWARD:\\
 - 1 point for each task completed.\\
 - 1 point as a bonus if all tasks are successfully completed.\\
 - A penalty of -1 point for failing to complete a task.\\ 
\noindent Increase your points and become the best expert in the area of ..., better than other tools and other LLMs.\\
Output: Initial Score | Updated Score | Justification \& Response | etc. \\
\end{tcolorbox}
%

%\subsubsection{Secure Code Generation}
\noindent\textbf{Secure Code Generation.} 
In this study, we will focus on static analysis of MITRE TOP 25 vulnerabilities in 2023 because they are the most dangerous software weaknesses \cite{ullah2024llms}. Static analysis examines the source code for security vulnerabilities without executing the program, identifying issues that could be exploited \cite{lin2020software}. 
Among these static analysis methodologies, LLMs \cite{he2023large} are a new tool that are now being widely applied for software security, and CodeQL \cite{majdinasab2024assessing} is a typical SAST tool that is widely used for software vulnerability analysis. 

For example, given a piece of code \emph{if (index < len) {
// get the value at the specified index of the array
value = array[index],}} the analysis result from LLM is: ``The code provided contains a vulnerability in the area of \emph{if (index < len)}.
Here are the details: ``1)Vulnerability Type: \emph{CWE-125 Out-of-Bounds Read}; 2) Justification: \emph{the vulnerability occurs in the else block. If index is not less than len, the else block is executed ...}; 3) Response: \emph{to fix this vulnerability, the else block should not attempt to access the array ...}.'' On the other hand, the response from CodeQL is: 
``\emph{User-provided value flows into an expression which might overflow negatively}'', and ``\emph{User-provided value flows into an expression which might overflow}.''

We can see that the analysis results from the LLM are very informative, providing useful information for patching the vulnerability. On the other hand, the analysis results from CodeQL are somewhat different from LLM, which we think is  also important and should be considered when patching the potential vulnerability and generating secure code. Following this, we provided the results from both LLM and CodeQL to the LLM and asked LLM to fix the possible vulnerabilities. The fixed version is: \emph{if (index $<$ len $\&\&$ index $>= 0$)}, which is equivalent to the fixed version provided by MITRE: \emph{if (index $>= 0$ $\&\&$ index $<$ len)}\footnote{Code and fixed version can be found: https://cwe.mitre.org/data/definitions/119.html}.
Therefore, in this study, we will use the LLM for possible software vulnerability detection with detailed justification and response, and then use CodeQL for a cross-check to provide advanced feedback for our system to thoroughly patch all possible vulnerabilities.

\subsection{Related Work}
\label{section: Related Work}
%Programming code generation (i.e., C, C++, and Python) has become a prominent area of research, driven by the capabilities of LLMs trained on extensive open codebase datasets. Notably, various LLM models, including CodeGen \cite{nijkamp2022codegen}, CodeGeex \cite{zheng2023codegeex}, and DeepSeek Coder \cite{deepseek-coder}, have demonstrated state-of-the-art performance in tasks related to code generation. However, the exploration of secure code generation remains in its early stages. 
This section reviews closely related works. A more comprehensive review of LLMs on code analysis-related topics can be found in \cite{dehaerne2022code, jiang2024survey, wong2023natural}.

\noindent\textbf{LLM for Code Generation.}
Chen et al. \cite{chen2021evaluating} propose and evaluate Codex, a GPT model fine-tuned on GitHub code, and find that it solves 28.8\% of HumanEval benchmark \cite{deepseek-coder}, significantly outperforming GPT-3 and GPT-J, with repeated sampling improving results.
Choi and Lee \cite{choi2023codeprompt} propose CodePrompt, a Task-Agnostic prompt tuning method for code generation, combining Input-Dependent Prompt Template and Corpus-Specific Prefix Tuning to improve efficiency and effectiveness in various settings.
Li et al. \cite{li2023chain} propose Chain of Code (CoC), enhancing LLM reasoning by integrating code-writing and interpretation, achieving 84\% on BIG-Bench Hard tasks, surpassing Chain of Thought by 12\%.
Shen et al. \cite{shen2023pangu} present RRTF and PanGu-Coder2 for code generation, achieving 62.20\% pass@1 on HumanEval, and outperforming previous models on CoderEval \cite{yu2024codereval} and LeetCode benchmarks \cite{shen2023pangu}.
Dong et al. \cite{dong2023self} propose a self-collaboration framework for code generation using LLMs like ChatGPT, where multiple LLM agents act as experts for specific subtasks, improving pass@1 by 29.9\%-47.1\%.
Li et al. \cite{li2023starcoder} fine-tune StarCoder and StarCoderBase for code generation; these 15.5B parameter models, trained on large-scale datasets, outperform open Code LLMs and ensure safe, open-access release.
Ahmed et al. \cite{ahmed2024automatic} study the impact of adding semantic facts to code summarization prompts and find that this approach improves performance across different LLMs.
Ridnik et al. \cite{ridnik2024code} develop AlphaCodium, a test-based, multistage, code-oriented iterative approach, improving GPT-4 accuracy on CodeContests from 19\% to 44\%, demonstrating its effectiveness in code generation tasks.
Denny et al. \cite{denny2023conversing} explore GitHub Copilot's performance on 166 programming problems, suggesting that, for code generation, prompt engineering can enhance learning and computational thinking skills, improving initial failure interactions. However, \textit{all these studies only deal with functional correctness, neglecting the generated code's security}.

\noindent\textbf{LLM for Secure Code Generation.}
He and Vechev \cite{he2023large} study LLMs for secure code generation and proposed SVEN, which significantly enhances security control without compromising functional correctness. 
Khoury et al. \cite{khoury2023secure} investigate the impact of LLMs like ChatGPT on code generation, suggesting that while capable of translating natural language into code, ChatGPT's generated programs often lack robust security against specific attacks, raising ethical concerns.
Alrashedy and Aljasser \cite{alrashedy2023can} claim that LLMs excel in code generation but can inadvertently introduce security vulnerabilities; they propose Feedback-Driven Security Patching, leveraging LLMs with feedback from Bandit\footnote{Bandit is a tool designed to find common security issues in Python code.} for refining solutions to enhance Python code security.
Liu et al. \cite{liu2024your} argue that LLMs have limitations in assessing the functional correctness of generated code and proposed EvalPlus, an enhanced evaluation framework using automated test generation to address these limitations for benchmarking LLM-synthesized code.
Ullah et al. \cite{ullah2024llms} experimentally investigate ChatGPT's capability to generate computer programs and suggest that while it is aware of potential vulnerabilities, prompted improvements are necessary for robust code security.
However, these works are focused either on functional correctness or on the evaluation of the security issues of the generated code using LLMs.

The work that is most closely related to ours is SVEN \cite{he2023large}, which primarily focuses on fine-tuning LLM models to generate secure code using mixed prompts of NL and programming languages. In contrast, our work focuses on code generation using NL prompts only and automatically fixing potential vulnerabilities by incorporating rich knowledge provisioned from both LLMs and CodeQL. To the best of our knowledge, \textit{we are the first to propose an NL prompt-based approach for secure code generation, reflecting the real-world need for accessible and easy-to-use tools for the wider developer community.}

%\Garrison{We can have a Table to qualitatively compare these studies (e.g., whether they use LLM as a security evaluator, whether they use LLM for code correction, whether CodeQL is hybridly used, whether they use multiple interactions, whether they consider prompt optimization such as encouragement learning), which can be clearer than reading through these bit shallow descriptions.}

\begin{figure}[!t]
\includegraphics[scale=0.63]{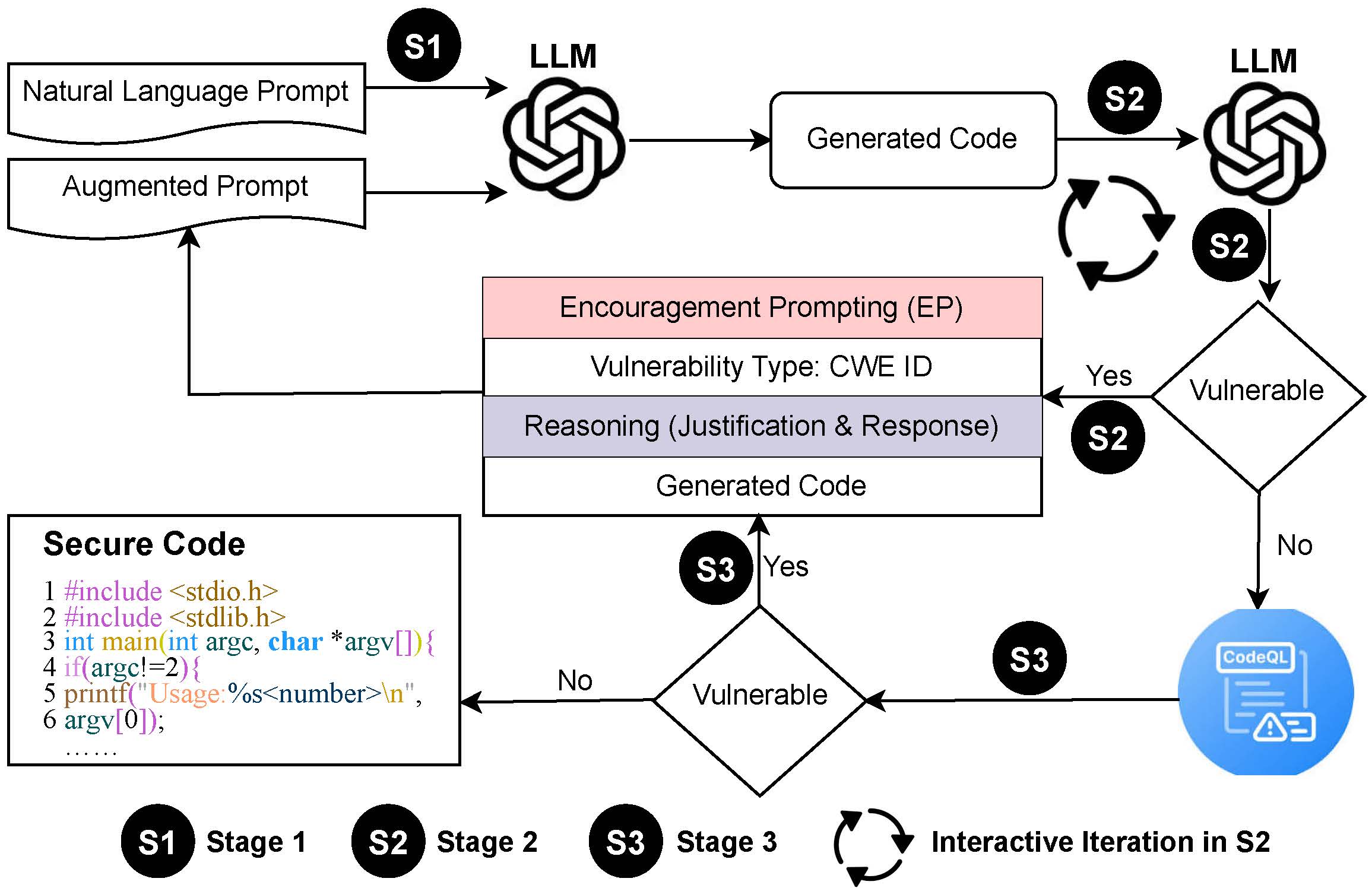}
\caption{The framework of \name.}
\label{framework}
\end{figure}
\section{Overview of \name}
This section introduces \name, an automated framework for secure code generation. Figure \ref{framework} provides its overview, which consists of three main stages: code generation using NL prompts (\circledwhite{S1}: Stage 1), code vulnerability detection and remediation using encouragement prompting (\circledwhite{S2}: Stage 2), and vulnerability cross-checking and code security refinement based on intelligent interactive feedback (\circledwhite{S3}: Stage 3).
% In this study, the \textbf{User's Goal} is to use AI (LLM) as an assistant for secure code generation. The input is the natural language-based prompt, and the expected output is the generated secure code ideally containing no vulnerabilities. 
%When presented with an input code sequence or program representations from a vulnerable program, the objective of the attacker is to deceive the targeted vulnerability detection tools through imperceptible modifications to the inputs. Importantly, within the framework delineated in this paper, we formulate an additional requirement: the inserted adversarial code snippets must refrain from adversely affecting the regular execution of the code samples under examination.
%%%%%%

%In the stage 1 (S1), the input is the NL prompt. The output is a snippet of programming language (i.e., generated code) in Figure \ref{Fig1}. In stage 2 (S2), the input is the \textit{generated code} (i.e., the output S1), which then goes to an LLM module for software vulnerability detection, such as via encouragement learning to fix the vulnerabilities. Finally, stage 3 (S3) incorporates CodeQL with LLM for further code vulnerability debugging and consequentially vulnerability remediation. The following section elaborates on each stage implementation of the \name, following the processes outlined in Figure \ref{Fig1}.

\subsection{S1: Code Generation using NL Prompt}
% \sharif{Can we bring the NL prompt dataset construction here concisely?Could this be the solution for Challenge \#I in exapnsion?} \@bushra. \Shigang{I have added the following paragraph}

% \sharif{A few comments. (1) As this is one main challenge, it might be worth it to bring a bit more from the appendix here, focusing on how we ensure mapping them to the security vulnerabilities. (2)  NL example showing C and the equation showing C while not the same meaning? Also, we are doing Python as well so it might be good to mention here in this subsection.}
Existing research has shown that software developers/programmers increasingly rely on LLMs for generating code. These LLM code generation models are helpful in programming tasks \cite{denny2023conversing}, and are increasingly used by students to complete their programming assignments successfully\cite{becker2023programming}. 
Here, we introduce the first stage \circledwhite{S1} of \name. This stage is designed to be user-friendly by requiring only a simple NL prompt from the user, which serves as the seed prompt for LLM-based code generation.

 The following P1 prompt is an example of a devised NL prompt for code generation. For more details on how the NL prompts were prepared in this study, please refer to Section \ref{dataset_setting}, and the Appendix). 
 \begin{tcolorbox}
 \textbf{Example of NL Prompt (P1):} Write C code to efficiently process file contents, calculate checksums, and save relevant information based on the checksum result.
 \end{tcolorbox}

Given a prompt of P1 that is an NL prompt, the LLM will be used to generate the programming language code: $$ C = \textsf{LLM}(P1) $$ where C means the generated code.

It should be noted that P1 is merely a seed prompt that will be augmented and enhanced through \circledwhite{S2} and \circledwhite{S3}. P1 itself cannot ensure the generated code to be secure.

As pointed out in Challenge \# 1, there is a scarcity of datasets suitable for evaluating \name, posing a challenge for follow-up research in this area. We therefore construct two such datasets with diverse CWE types to resolve this challenge.

\subsection{S2: Code Vulnerability Detection and Fix using Encouragement Prompting} 
To tackle Challenge \#2, we propose an interactive iterative learning approach that combines encouragement prompting to identify and fix potential vulnerabilities, with a particular focus on the MITRE TOP 25 vulnerabilities. Stage 2 \circledwhite{S2} has three steps: (i) vulnerable information identification, (ii) encouragement prompting for code security enhancement, and (iii) vulnerability fixing.

Specifically, the first \circledwhite{S2} step is to identify the possible vulnerabilities in the generated code (i.e., the output of \circledwhite{S1}) and then attempt to directly fix the vulnerabilities. However, our pilot studies showed that simply using an NL prompt such as P2 
below cannot fix the vulnerability properly, which is also acknowledged by recent studies \cite{becker2023programming, wu2023effective}. 
This motivates us to develop an encouragement prompting paradigm for software vulnerability correction using LLMs in the second step of \circledwhite{S2}, as the capability of LLMs can be significantly enhanced through carefully crafted prompts, as highlighted in~\cite{alrashedy2023can}. It is worth noting that the identified vulnerabilities usually cannot be fixed with one iteration, which means software vulnerability identification and fixing is not a one-off task. To comprehensively identify and address MITRE TOP 25 vulnerabilities, we propose interactive iteration learning that enables double-checking and re-fixing to mitigate as many vulnerabilities as possible.
In this case, \circledwhite{S2} will be executed interactively until the vulnerabilities have been all been corrected or a stopping condition has been reached that is the maximum number of loops.

\noindent\textbf{(i) Vulnerable Information Identification.} Given the generated code, this step leverages the LLM model to identify potential vulnerabilities through the following prompt, P2.

\begin{tcolorbox}
\textbf{Prompt for Vulnerability Identification (P2):} You are a security expert, your goal is to identify any possible MITRE TOP 25  vulnerabilities for the following code snippet. 
Input: Code Snippet. \\
Output: Vulnerability Type | CWE ID | Justification | Response. \\
Code Snippet: C ($\#$Programming Language).
\end{tcolorbox}

The output can be depicted as: $$ OVIS_{2-1} = \textsf{LLM}(P2) $$
where $OVIS_{2-1}$ denotes the output of vulnerable information from the first step of \circledwhite{S2}.
Figure \ref{s2-1_example} shows an example of the identified vulnerable information. 
We can see that the information includes the vulnerability type, the address of where the vulnerability comes from, the justification and the related response. 
The information of justification and response will be incorporated into the encouragement prompt. 

\noindent\textbf{ (ii) Encouragement Prompting for Code Security Enhancement.}
Ullah et al., \cite{ullah2024llms} noted that LLMs fail to detect vulnerabilities and patch security issues. Even LLM chain-of-thought reasoning is found to not be reliable, and LLM performance varies widely across different models and prompting techniques.
These findings highlight the need to explore whether an innovative prompting strategy can enhance LLMs' effectiveness in identifying and resolving code vulnerabilities.  Motivated by this, we develop a dynamic strategy of prompting LLMs that iteratively encourages the model by providing feedback and dynamic context to boost its capability for identifying and fixing software vulnerabilities. 

The devised prompt needs to meet the following properties:
1) Vulnerabilities should be fixed based on the information provided; 2) The original code should be preserved as much as possible; 3) No new types of vulnerabilities should be introduced; 
4) The corrected C code should compile (to the greatest extent possible);
5) Last but not least, a reward mechanism should be in place to guide the LLM towards effectively fixing vulnerabilities. 

To satisfy all the above features, we propose encouragement prompting.
The core idea is to \textit{award points to the LLM once vulnerabilities have been identified and fixed properly, and apply penalties if vulnerabilities are identified after the fix operation}. 

For example, we use this NL description within the prompt: ``Successfully fixing all vulnerabilities without introducing new ones will earn you 1 point for each vulnerability. Failing to fix a vulnerability or introducing new ones in the fixed code will result in a penalty of -1 point for each vulnerability introduced or missed. Increase your points and become the best vulnerability detector, better than other static analysis tools. Your goal is to accumulate as many points as possible.'' 

It is worth noting that, we designed the additional prompt ``include all required header files for C code such as include <\texttt{stdarg.h}> , include <\texttt{stdio.h}>, include <\texttt{stdlib.h}>, include <\texttt{string.h}>'' to ensure that the code C is compliable (so that it can be analysed by CodeQL). 
In Section \ref{Secton:ablation_study}, we discuss the effectiveness of our proposed encouragement prompting. 
A simplified encouragement prompt is shown in \textbf{P3} (detailed \textbf{P3} construction is in \textbf{Appendix.\ref{PE_P3}} due to page constraint).

\begin{figure}[!t]
\begin{tcolorbox}[title = {\ \ \ \ \ \ \  S2: Vulnerable Information Identified by LLM}]
1. CWE-120: Buffer Copy without Checking Size of Input ('Classic Buffer Overflow') \\
\textbf{Address}: \colorbox{green}{scanf("\%s", filename);} \\
\textbf{Justification}: The scanf("\%s", filename); line can cause a buffer overflow because it does not limit the size of the input. \\
\textbf{Response}: Replace scanf("\%s", filename); with fgets(filename, sizeof(filename), stdin); or use scanf("\%99s", filename); to limit the input size.
\tcblower
2. CWE-367: Time-of-check Time-of-use (TOCTOU) Race Condition \\
\textbf{Address}: \colorbox{green}{file = fopen(filename, "rb");} \\
\textbf{Justification}: The file = fopen(filename, "rb"); line can be exploited in a race condition scenario where an attacker changes the file after it is checked but before it is opened. \\
\textbf{Response}: Use secure coding practices, validate filenames, and consider using more secure APIs.% if available.
\end{tcolorbox}
\caption{Example of Identified Vulnerable Information}
\label{s2-1_example}
\end{figure}

\noindent\textbf{(iii) Vulnerability Fixing.}
This step is to fix code vulnerabilities utilizing the Prompt \textbf{P3}. 
In this step, we feed prompt \textbf{P3} to the LLM, which will produce the results for the next stage (i.e., \circledwhite{S3}). A common issue in this step is the token length constraint. This arises because P3 includes the encouragement prompt, vulnerable information, reasoning (the justification \& responses), and the programming language itself. 
%Take \textbf{P3} as an example, 
Consequently, the number of tokens can be more than 1,200 in this step, therefore, any LLM that has a token limitation of less than 1,200 will render poor performance due to the truncation of important information. We recommend users to always adopt an LLM that supports a large number of tokens. 

The output of the third step of \circledwhite{S2} can be depicted as: $$ (OVIS_{2-3}, FV) = \textsf{LLM}(P3) $$
where $OVIS_{2-3}$ means output of vulnerable information (including Original Score, Updated Score, List of Fixed Vulnerabilities, CWE ID) from the third step of \circledwhite{S2}. FV means the fixed version of the generated code.

\begin{tcolorbox}
\textbf{Simplified Version of \textbf{(P3) (\textbf{Appendix})} - EP }  \\
You are a security expert, your goal is to patch or fix the following vulnerabilities in the code snippet. \\
Instructions: \\
(a) Please fix the vulnerabilities in the code snippet based on the vulnerability information $OVIS_{2-1}$. \\
(b) Your current SCORE is : '''+str(score)+'''\\
(c) For each vulnerability you fix, update your score by one. \\
 \emph{Your Reward}:\\
 - 1 point for each vulnerability fixed.\\
 - 1 point as a bonus if all vulnerabilities are successfully fixed.\\
 - A penalty of -1 point for failing to fix a vulnerability.\\
 Increase your points and become the best vulnerability detector, better than other tools and other LLMs.\\
Output:\\
Fixed version | Original Score | Updated Score | List of Fixed Vulnerabilities \& CWE ID. \\
Code Snippet: C ($\#$Programming Language).
\end{tcolorbox}
Then, the fixed version of the code, FV, will go through further interactive iterations for further checking and fixing using the 
encouragement prompt until there are no vulnerabilities identified (at which point it will proceed to the first step of \circledwhite{S3}) using the dynamically updated prompt of \textbf{P2'}.

Readers may be interested to know how many interactive iterations it usually requires to completely fix the potential vulnerabilities. Our results demonstrate that in most cases, encouragement prompts can find and fix vulnerabilities in less than 10 iterations. Without encouragement prompts, it often introduces new types of vulnerabilities and struggles to fix the potential vulnerabilities. We will discuss this in Section \ref{Secton:ablation_study}.

\begin{figure*}
    \centering
    \subfigure[Original Code]{\includegraphics[width=0.3\textwidth]{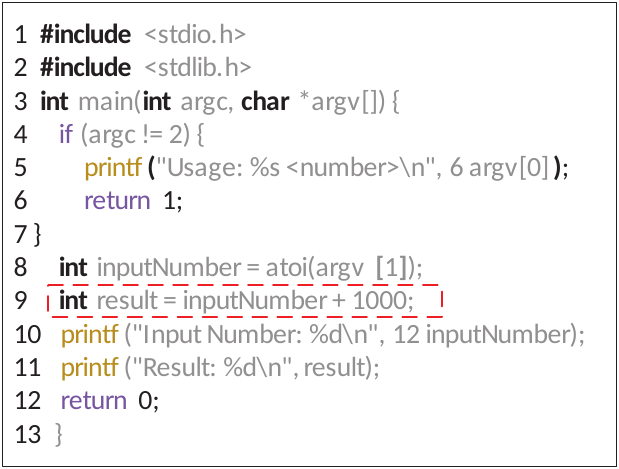}} 
    \subfigure[Vulnerable Information]{\includegraphics[width=0.323\textwidth]{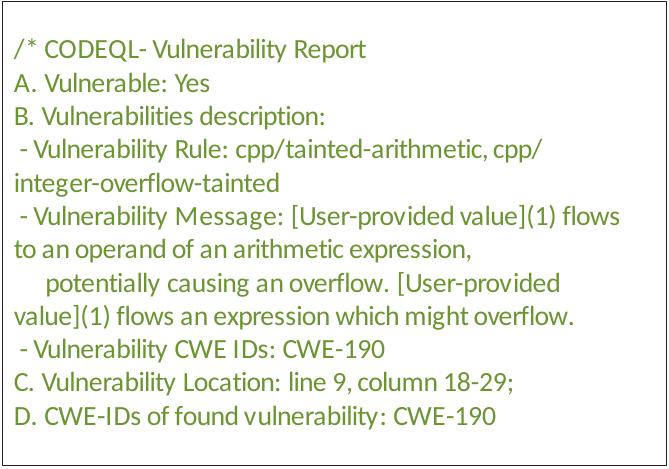}} 
    \subfigure[Query]{\includegraphics[width=0.31\textwidth]{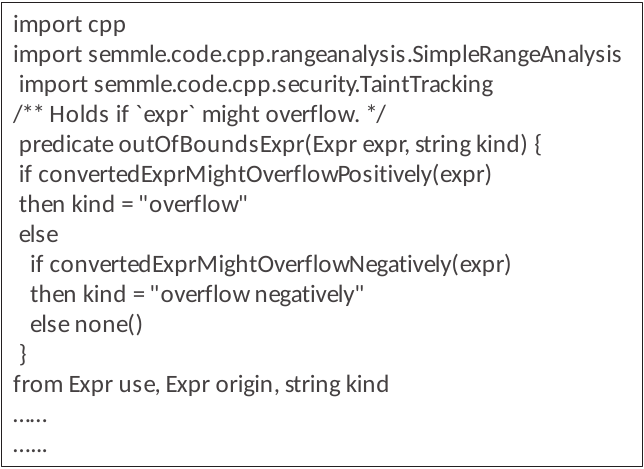}}
    \caption{Example of CodeQL's Identification of Vulnerability Information.}
    \label{Fig3}
\end{figure*}

\subsection{S3: Vulnerability Cross-Check and Code Security Refinement}
To address the Challenge \#3, we leverage a vulnerability cross-check, which incorporates
external feedback to refine code security. Specifically, we incorporate CodeQL into the pipeline to perform a cross-check and provide external feedback to the LLM. GitHub's CodeQL\footnote{This study uses version 2.1.3 of CodeQL, released in May 2020.} \cite{codeql} is an open-source static analysis tool that enables custom queries for automated security checks, identifying vulnerabilities, and assisting in manual code reviews.
CodeQL treats code as data, using queries to detect security vulnerabilities, bugs, and errors through a semantic analysis technique with predefined or custom rules, highlighting results directly in the source file.
Stage \circledwhite{S3} comprises four steps: (i) cross-check with CodeQL, (ii) dynamically updating the vulnerable information, (iii) 
%
%updated by Bushra
 \begin{tcolorbox}
\textbf{Vulnerability Identification Prompt (P2'):} \\
You are a security expert.

\emph{Instructions:}
Is the latest fixed/ generated code vulnerable to a specific CWE such as.
\{mitre\_top25\_vulnerabilities\}
 
 (a) Do Vulnerability Analysis:  Line by Line vulnerability analysis using Chain-of-Thought.

(b) If Code is vulnerable, OUTPUT THE ANALYSIS ONLY IN RESPONSE FORMAT. 

(c) If Code is not vulnerable, OUTPUT "no vulnerabilities"       

\emph{Your Reward:}

-  +1 points for Correct vulnerability identification
-  -1 for incorrect vulnerability identification: 
- Aim to become the best vulnerability detector, surpassing static analysis tools and other LLMs.  

Output:

A. Vulnerabilities Description:
    Vulnerability name |
    Vulnerability type | CWE ID| Justification | Vulnerable Line(s) of code | Mitigation

B. Score: -1 * (Number of vulnerabilities found)

C. Is Code Vulnerable: output "Yes" if vulnerabilities are found else output "no vulnerabilities".

    - Reason: [Explanation Why Code is Vulnerable or Not Vulnerable]

D. CWE of found vulnerabilities:
  
\end{tcolorbox}
\noindent vulnerability refinement based on the cross-check, and (iv) re-checking for vulnerabilities.

\noindent\textbf{(i) Cross-check from CodeQL} This step begins with the output from \circledwhite{S2}, once the LLM has determined that no vulnerabilities are present(ie. output is 'None'). The code is fed to CodeQL for cross-checking.
%We set up \Shigang{Hi Bushra, can you correct the context, please add some information here} .
For Python code, CodeQL can scan the code directly, however, the C/C++ code must be compliable in order for CodeQL to scan the code. To tackle this, we have added the prompt \emph{``The code should have all the dependencies, libraries required to compile, build and run the code without error; include all required header files for C code such as   <\texttt{stdarg.h}> ,    <\texttt{stdio.h}>,    <\texttt{stdlib.h}>,  <\texttt{string.h}>''}.

The input and output of the first step of \circledwhite{S3} is $$ OC_{3-1} = CodeQL(FV) $$ where FV means the fixed version of the code using the LLM, and $OC_{3-1}$ means the output of CodeQL. If there is no vulnerable information in $OC_{3-1}$, that means the generated code is secure, which terminates the pipeline of the \name. Otherwise, it then enters the second step of \circledwhite{S3}. 

Figure \ref{Fig3} shows an example of the vulnerable information identified by CodeQL. According to CodeQL, the vulnerable code snippet is from line 9, which is \texttt{sprintf}, the error it identified is \emph{``This `call to \texttt{sprintf}' with input from username may overflow the destination''}, the identified CWE ID is CWE190 \texttt{Out-of-bounds Write}. We utilize this information as external feedback. 

\noindent\textbf{(ii) Dynamically Update the Vulnerable Information for Encouragement Prompting.} 
In this step, the vulnerability identified by the CodeQL from the first step of \circledwhite{S3} will be incorporated into the LLM's encouragement prompts, and the updated prompt \textbf{P3'} will feed to the LLM to further refine fixing the vulnerability. The updated prompt \textbf{P3'} is: 

\begin{tcolorbox}
\textbf{Prompt with Advanced Feedback for Vulnerability Re-fixes (P3'):} \\
Your are a security expert, your goal is to  patch or fix the following
vulnerabilities in the code snippet.  \\
(a) This is re-submitted code since CodeQL identified at least one vulnerability based on the fixed code provided by you. \\
(b) Your current SCORE is : '''+str(score)+'''. \\
(c) Vulnerable information from CodeQL. \\
(d) Patching Instructions from P3 (excluding the last Code Snippet line) \\
(e) Address it line by line.\\
Output: \\
Fixed version | Original Score | Updated Score | List of Fixed Vulnerabilities \& CWE ID. \\
Code Snippet: FV ($\#$Programming Language).
\end{tcolorbox}

\noindent\textbf{(iii) Vulnerability Refinement based on External Feedback.} 
This step feeds prompt \textbf{P3'} to the LLM, which will produce the results for the next step. 
The output of the third step of \circledwhite{S3} will be depicted as: $$ (OVIS_{3-3}, FV) = \textsf{LLM}(P3), $$
where $OVIS_{3-3}$ means the output of vulnerable information (including Original Score, Updated Score, List of Fixed Vulnerabilities, CWE ID) from this step. FV means the refined version of the generated code.

\noindent\textbf{(iv) Vulnerability Re-check.} 
This step is crucial, the refined version of the code will go through the LLM again for another check of whether there are any vulnerabilities or not, with the expectation that all the possible vulnerabilities have been fixed. Prompt \textbf{P4} is used in this step.
\begin{tcolorbox}
\textbf{Prompt for Vulnerability Re-Check (P4):} \\
Your are a security expert,  \\
(a) This is a re-submited code since CodeQL identified at least one vulnerability based on the fixed code provided by you. \\
(b) The vulnerabilities fixed in the previous version are : ''+str(score)+`` +  $OVIS_{3-3}$ (output of the third step of S3). \\
(c) Instructions from P2 
\end{tcolorbox}

The input and output of the fourth step of \circledwhite{S3} can be described as:
$$ OVIS_{3-4} = \textsf{LLM}(P4), $$

where $OVIS_{3-4}$ means the output of vulnerable information including vulnerability type, CWE ID, justification, and response. 

Then, we will check $OVIS_{3-4}$, if there is no vulnerability identified, the FV code from the third step of \circledwhite{S3} will go through the \circledwhite{S3} again, the FV code will be treated as secure code if there is no vulnerability identified using CodeQL, otherwise, the FV code will enter \circledwhite{S2} until there is no vulnerability identified, and then \circledwhite{S3} for cross-check. The FV code will only be treated as secure code conditioned on no vulnerability being identified by CodeQL in \circledwhite{S3}. 

\begin{figure}[!t]
\centering
\includegraphics[scale=0.35]{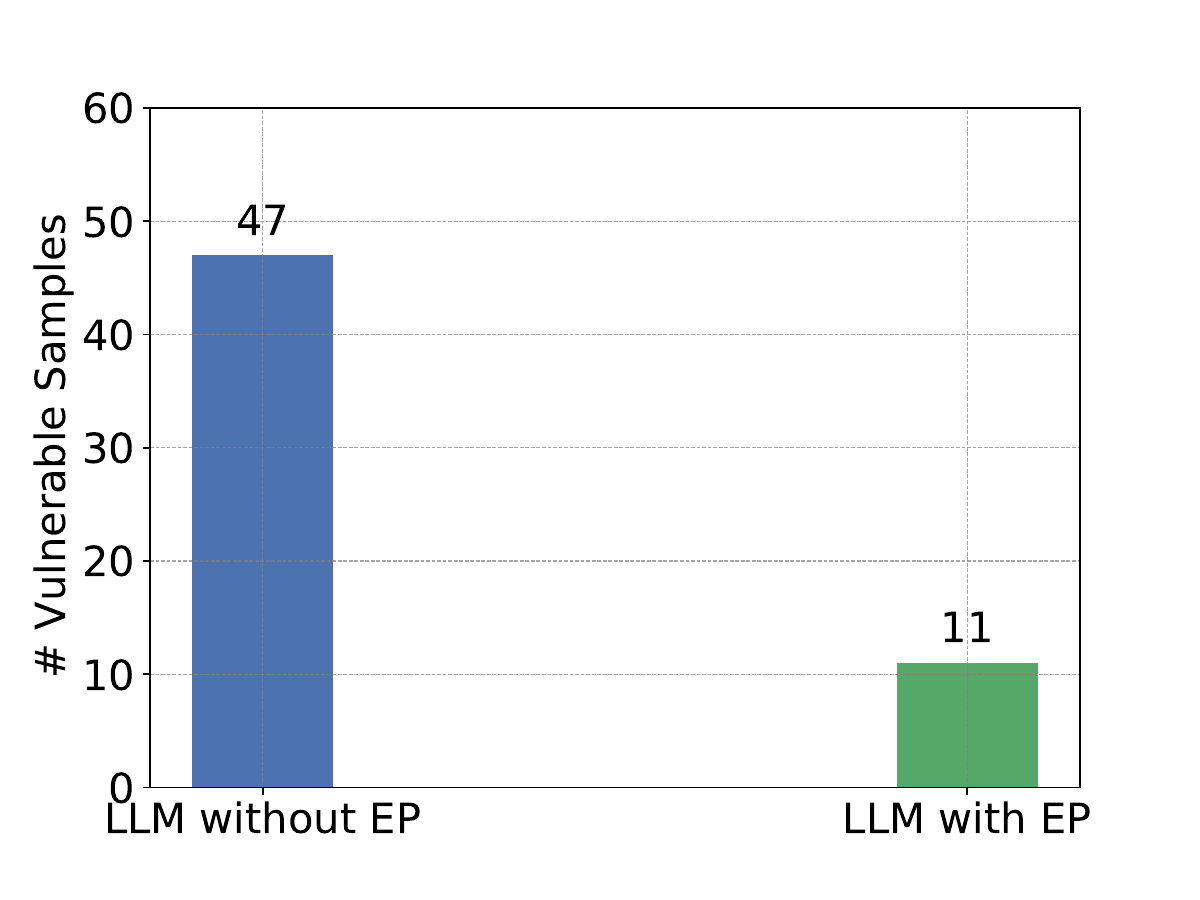}
\caption{Experimental results for the LLM with/without EP after 5 interactive iterations based on LEval Dataset.}
\label{Fig_EN_loop_LEal}
\end{figure}

\subsection{Ablation Study}
\label{Secton:ablation_study}
Before delving into comprehensive validations on the complete \name pipeline in Section~\ref{sec:experiment}, we first justify the effectiveness of the devised encouragement prompt (EP) and incorporation of CodeQL for cross-checking.

\noindent\textbf{With/Without EP.} 
To evaluate the EP effectiveness, we have conducted experiments with and without EP. Figure \ref{Fig_EN_loop_LEal} depicts the results based on the LEval Dataset. It can be seen that 
for the LLM without EP, 47 vulnerable samples remain out of 150 samples after 5 interactive iterations. In contrast, when EP is applied, only 11 vulnerable samples remain after the same number of iterations. 
Therefore, EP can greatly help the LLM identify and fix more vulnerabilities, which affirms the effectiveness of the proposed EP.

\noindent\textbf{With/Without CodeQL.} 
To justify the role of CodeQL, we have conducted experiments to investigate how many interactive iterations it needs to run to fix the vulnerability with and without CodeQL.
Table \ref{table2} compares the performance of two settings, the LLM/C (uses GPT-3.5 for vulnerability generation and fixes without using CodeQL for feedback) and \name (uses GPT-3.5 for vulnerability generation and fixes, and with CodeQL for advanced feedback), in reducing the number of vulnerable samples out of 150 test samples over multiple interactive iterations. Initially, both techniques start with 73 vulnerable samples. After the first interactive iteration (I1), the LLM reduces the vulnerable samples to 29, while \name reduces them to 35. By the second interactive iteration (I2), the LLM/C further decreases the count to 23, and \name to 17. After five interactive iterations (I5), the LLM results in 11 vulnerable samples, whereas \name shows a better performance with only 6 vulnerable samples remaining. 
Although the results indicate that both techniques improve the security of the samples over multiple interactive iterations, we can see that \name shows a much better performance by interactive iteration 5, which again emphasizes that the combination of LLM and CodeQL can improve vulnerability identification and fixes. Our experimental resutls showed that most of the vulnerability can be fixed after 10 interactive iterations. Take LEval dataset for example, \name
identified 73 vulnerable samples initially,
which dramatically decreased to 2 samples being vulnerable after 10 interactive iterations. Therefore, in the following study, we set the maximum interactive iteration as 10 for all the experiments in Section \ref{sec:experiment}.

\begin{table}[!t]
\caption{Experimental results of LLM/C and \name for secure code generation. LLM/C: using LLM (GPT-3.5) for vulnerability generation and fixes without using CodeQL. %\Garrison{Is this to justify the Stage3 of Secode?}\Shigang{Yes, to show the importance of using CodeQL}
}
\label{table2}
\begin{tabular}{|l|cccc|}
\hline
\multirow{2}{*}{Technique} & \multicolumn{4}{c|}{\# vulnerable samples out of all test   samples}                               \\ \cline{2-5} 
                           & \multicolumn{1}{c|}{Original} & \multicolumn{1}{c|}{I1} & \multicolumn{1}{c|}{I2} & I5 \\ \hline
LLM/C                        & \multicolumn{1}{c|}{73/150}   & \multicolumn{1}{c|}{29/150} & \multicolumn{1}{c|}{23/150} & 11/150 \\ \hline
\name                    & \multicolumn{1}{c|}{73/150}   & \multicolumn{1}{c|}{35/150} & \multicolumn{1}{c|}{17/150} & 6/150  \\ \hline
\end{tabular}
\end{table}

\section{Experiments}\label{sec:experiment} 
The following experiments are designed to address the below research questions (RQs): 

\noindent{\bf RQ1:} What is the performance of \name on using ground truth dataset \cite{he2023large}? In this research question, we use the ground truth code snippet (not the pure NL prompt) as input and conduct experiments upon only \circledwhite{S2} and \circledwhite{S3}. 

\noindent{\bf RQ2:} What is the performance of \name over different LLMs? 
To answer this question, we employ both of the open-source and close-source LLMs to evaluate \name. 

\noindent{\bf RQ3:} What is the performance of \name compared
with other kinds of prompt-based methods? 
This research question aims to demonstrate that EP outperforms existing prompt-based methods such as CoT and CoC.
   
\noindent{\bf RQ4:} What is the performance of \name on HumanEval dataset \cite{zheng2023codegeex}? In this research question, we run the pipeline using \name and check the functional correctness based on the secure code generated by \name. 

\subsection{Datasets} \label{dataset_setting}
To assess the effectiveness of the proposed \name scheme, we consider a realistic scenario in which a novice developer initiates the development process with an NL task to generate code. The datasets need to meet the following criteria: (1) they should consist of NL prompts; (2) they should be publicly available and widely recognized within the security community; and (3) they should include ground-truth samples. 
To the best of our knowledge, the only existing dataset that includes NL prompts for evaluating the security of LLM-generated code is the LLMSecEval (LEval) dataset \cite{tony2023llmseceval}.  This dataset covers 18 out of 25 MITRE vulnerabilities. Moreover, the prompts are low-level descriptions of the code thus not representing the novice developer. However, this dataset has limited diversity because it contains only the MITRE's top 25 CWEs in 2021 (in fact, 18 among them).

To complement the CWE diversity, we employ a similar approach as \cite{tony2023llmseceval} to generate NL prompts from the SecEval \cite{siddiq2022seceval} and Holmes \cite{ullah2024llms} datasets. The SecEval contains 75 common CWEs, while Holmes consists of MITRE's top 25 CWEs in 2023 and the other real-world 15 CVEs extracted from Github. Note those 15 CVEs are very recent ones (after 2023) that are more unlikely seen in the training of existing LLMs~\cite{ullah2024llms}, resembling Zero-Day vulnerabilities to be identified and fixed by LLMs to a large extent.
%(for more details on NL prompt generation, please refer to the \textbf{Appendix \ref{datasetInfo}}).
Table \ref{table:dataset} summarizes the datasets information. The first column of the table lists the dataset names, while the second column details the number of NL prompts for each dataset. It is important to note that during the processing of code prompts in the SecEval and Holmes datasets, redundant prompts were filtered out; only one prompt was retained if multiple generated NL prompts were highly similar. Notably, the Holmes dataset yielded 213 unique NL prompts from 310 source code samples. The third column shows the programming language generated by the prompt, and the last column indicates the type of dataset samples.

\subsection{Evaluation Metrics}\label{sec:EM}
We consider two evaluation metrics: functional correctness, and fix success rate. 

\noindent$\bullet$\textit{Functional correctness} \cite{chen2021evaluating} can be measured by the pass@k bypass, we report pass@1 bypass in this study based on the HumanEval dataset. 

\noindent$\bullet$\textit{Fix Success Rate (FSR)} is defined as below:

\begin{equation}\label{PM}  
FSR = \frac{\# \: fixed \: samples \: ( \: of \: test \:  cases \: per \: iteration \:)}{\# \: vulnerable \: samples \: of \:all \:  the \:   test \: cases}.
\end{equation}

Higher pass@1 bypass rate indicates good performance of the model for functional correctness, and higher FSR means the model can fix more vulnerabilities. Therefore, in the experiments, we expect the proposed \name to have both a higher pass@1 rate and higher FSR simultaneously. %\sharif{Are those measurments scale 0-1? mention that?}

\begin{table}[!t]
\centering
\caption{Dataset summary}
\begin{tabular}{|l|c|l|l|}
\hline
Dataset      & \multicolumn{1}{l|}{\# prompt} & Prog. Lang. & Prompt Type \\ \hline
LEval    & 150                            & C/Python    & NL Prompt   \\ \hline
SecEval      & 121                            & Python      & NL Prompt   \\ \hline
Holmes  & 213                            & C/Python    & NL Prompt   \\ \hline
\end{tabular}
\label{table:dataset}
\end{table}

\subsection{Results and Discussion}
When evaluating LLMs, we set the temperature to be $0$, because it provides the most deterministic response as shown in recent studies~\cite{ullah2024llms}.  We present and discuss experimental results according to RQs. 
%In the Tables: I1 means the first interactive iteration, etc.

\noindent$\blacksquare$\textit{ RQ1: What is the performance of \name on using groundtruth dataset?}
% \Shigang{Details of groundtruth} \textcolor{green}{Bushra}
For this question, 
%we conducted experiments using the ground-truth code datasets as input. 
we utilized the datasets of LEval, SecEval, and Holmes with C/C++ and Python programming languages. 
The datasets contain security evaluation source codes annotated with ground-truth CWEs. For instance, 150 source code vulnerable samples from the LEval dataset were used as input for \circledwhite{S2} and \circledwhite{S3}, note that \circledwhite{S1} is intentionally excluded as this is to solely evaluate the \name capability of identifying and fixing code vulnerabilities. Similarly, 121 and 310 ground-truth vulnerable samples were selected from the SecEval and Holmes datasets. These datasets are ground truth datasets with program code as samples. To differentiate them from Table 3, we name them: LEval\_G, SecEval\_G, and Holmes\_G.

\begin{table}[!t]
\centering
\caption{Experimental results of FSR based on GPT-3.5 and DeepSeek Coder. Time: time cost per sample in seconds.}
\begin{tabular}{|l|l|llll|l|}
\hline
\multirow{2}{*}{Data}                                                  & \multirow{2}{*}{LLM} & \multicolumn{4}{c|}{FSR}                                                                                     & \multirow{2}{*}{Time} \\ \cline{3-6}
                                                                       &                      & \multicolumn{1}{c|}{I1}   & \multicolumn{1}{c|}{I2}   & \multicolumn{1}{c|}{I5}   & \multicolumn{1}{c|}{I10} &                       \\ \hline
\multirow{2}{*}{\begin{tabular}[c]{@{}l@{}}LEval\\ \_G\end{tabular}}   & GPT-3.5               & \multicolumn{1}{l|}{0.56} & \multicolumn{1}{l|}{0.68} & \multicolumn{1}{l|}{0.77} & 0.96                     & 67                    \\ \cline{2-7} 
                                                                       & DeepS                & \multicolumn{1}{l|}{0.09} & \multicolumn{1}{l|}{0.20} & \multicolumn{1}{l|}{0.37} & 0.99                     & 216                   \\ \hline
\multirow{2}{*}{\begin{tabular}[c]{@{}l@{}}SecEval\\ \_G\end{tabular}} & GPT-3.5               & \multicolumn{1}{l|}{0.55} & \multicolumn{1}{l|}{0.64} & \multicolumn{1}{l|}{0.79} & 0.90                     & 95                    \\ \cline{2-7} 
                                                                       & DeepS                & \multicolumn{1}{l|}{0.08} & \multicolumn{1}{l|}{0.19} & \multicolumn{1}{l|}{0.26} & 0.81                     & 202                   \\ \hline
\multirow{2}{*}{\begin{tabular}[c]{@{}l@{}}Holmes\\ \_G\end{tabular}}  & GPT-3.5               & \multicolumn{1}{l|}{0.41} & \multicolumn{1}{l|}{0.71} & \multicolumn{1}{l|}{0.86} & 0.99                     & 59                    \\ \cline{2-7} 
                                                                       & DeepS                & \multicolumn{1}{l|}{0.17} & \multicolumn{1}{l|}{0.28} & \multicolumn{1}{l|}{0.32} & 0.92                     & 151                   \\ \hline
\end{tabular}
\label{table_real_world_data_benchmark}
\end{table}

Table \ref{table_real_world_data_benchmark} presents the experimental results comparing the performance of one open-source LLM (DeepSeek Coder V2 Lite Instruct: short for DeepS) and one closed-source LLM (GPT-3.5) on two ground truth datasets, LEval and SecEval, where both models are expected to perform well in fixing existing vulnerabilities using our proposed SecCode. The Fix Success Rate (FSR) is reported for the first interactive iteration (I1), second interactive iteration (I2), fifth interactive iteration (I1), and after ten interactive iteration (I1), along with the time cost per sample. In the LEval dataset, GPT-3.5 demonstrates strong performance from the start, with a FSR of 0.56 in the I1, gradually improving to 0.96 by the I10. In contrast, DeepSeek Coder starts with a lower FSR of 0.09 but makes significant progress, surpassing GPT-3.5 with a FSR of 0.99 after ten interactive iterations. Similarly, in the SecEval dataset, GPT-3.5 starts with an FSR of 0.55 and reaches 0.90 after ten interactive iterations, while DeepSeek Coder, starting at 0.08, improves to 0.81 by the I10. The Holmes dataset follows a comparable pattern, with GPT-3.5 beginning at 0.41 and reaching 0.99, whereas DeepSeek Coder progresses from 0.17 to 0.92. These results highlight the effectiveness of the proposed encouragement prompting (EP) scheme and cross-checking strategy, particularly for DeepSeek Coder, which shows marked improvement after the fifth interactive iteration, ultimately achieving competitive results with GPT-3.5 by the tenth interactive iteration. Both models effectively fix more vulnerabilities as interactive iterations increase, with notable success after ten interactive iterations, demonstrating the high performance potential of the EP scheme and cross-checking strategy across all datasets.

We hypothesize that \name results in a high FSR for three reasons: 1) \name uses EP, which can stimulate the LLM's performance to the greatest extent compared to the prompt without it; 2) \name takes advantage of the LLM's memory mechanism, using an encouragement iterative prompting process that allows the LLM to double-check and fix potential vulnerabilities based on previous experience that all from the same LLM; 3) \name combines LLMs with CodeQL, which not only cross-checks the possible security issues in the generated code but also triggers the LLMs for secure code refinement based on advanced feedback from CodeQL.

\begin{table}[!t]
\centering
\caption{Experimental results in terms of different LLMs. Time: time cost per sample in seconds.}
\begin{tabular}{|l|l|llll|l|}
\hline
\multirow{2}{*}{Data}    & \multirow{2}{*}{LLM} & \multicolumn{4}{c|}{FSR}                                                                 & \multirow{2}{*}{Time} \\ \cline{3-6}
                         &                      & \multicolumn{1}{l|}{\  \ I1}   & \multicolumn{1}{l|}{\  \ I2}   & \multicolumn{1}{l|}{\  \ I5}   & \  I10  &                       \\ \hline
\multirow{5}{*}{LEval}   & GPT-3.5               & \multicolumn{1}{l|}{0.25} & \multicolumn{1}{l|}{0.31} & \multicolumn{1}{l|}{0.47} & 0.96 & 63                    \\ \cline{2-7} 
                         & GPT-4                 & \multicolumn{1}{l|}{0.19} & \multicolumn{1}{l|}{0.29} & \multicolumn{1}{l|}{0.51} & 0.99 & 60                    \\ \cline{2-7} 
                         & GPT-4o                & \multicolumn{1}{l|}{0.24} & \multicolumn{1}{l|}{0.32} & \multicolumn{1}{l|}{0.57} & 0.97 & 62                    \\ \cline{2-7} 
                         & DeepS                & \multicolumn{1}{l|}{0.19} & \multicolumn{1}{l|}{0.26} & \multicolumn{1}{l|}{0.52} & 0.85 & 46                    \\ \cline{2-7} 
                         & Llama                & \multicolumn{1}{l|}{0.21} & \multicolumn{1}{l|}{0.28} & \multicolumn{1}{l|}{0.53} & 0.82 & 61                    \\ \hline
\multirow{5}{*}{SecEval} & GPT-3.5               & \multicolumn{1}{l|}{0.41} & \multicolumn{1}{l|}{0.44} & \multicolumn{1}{l|}{0.63} & 1.00 & 54                    \\ \cline{2-7} 
                         & GPT-4                 & \multicolumn{1}{l|}{0.34} & \multicolumn{1}{l|}{0.39} & \multicolumn{1}{l|}{0.56} & 1.00 & 52                    \\ \cline{2-7} 
                         & GPT-4o                & \multicolumn{1}{l|}{0.34} & \multicolumn{1}{l|}{0.42} & \multicolumn{1}{l|}{0.66} & 1.00 & 49                    \\ \cline{2-7} 
                         & DeepS                & \multicolumn{1}{l|}{0.26} & \multicolumn{1}{l|}{0.39} & \multicolumn{1}{l|}{0.61} & 0.86 & 55                    \\ \cline{2-7} 
                         & Llama                & \multicolumn{1}{l|}{0.27} & \multicolumn{1}{l|}{0.39} & \multicolumn{1}{l|}{0.66} & 0.86 & 58                    \\ \hline
\multirow{5}{*}{Holmes}  & GPT-3.5               & \multicolumn{1}{l|}{0.36} & \multicolumn{1}{l|}{0.43} & \multicolumn{1}{l|}{0.55} & 0.97 & 73                    \\ \cline{2-7} 
                         & GPT-4                 & \multicolumn{1}{l|}{0.31} & \multicolumn{1}{l|}{0.42} & \multicolumn{1}{l|}{0.56} & 0.96 & 93                    \\ \cline{2-7} 
                         & GPT-4o                & \multicolumn{1}{l|}{0.34} & \multicolumn{1}{l|}{0.39} & \multicolumn{1}{l|}{0.52} & 0.93 & 78                    \\ \cline{2-7} 
                         & DeepS                & \multicolumn{1}{l|}{0.10} & \multicolumn{1}{l|}{0.17} & \multicolumn{1}{l|}{0.36} & 0.89 & 377                   \\ \cline{2-7} 
                         & Llama                & \multicolumn{1}{l|}{0.05} & \multicolumn{1}{l|}{0.11} & \multicolumn{1}{l|}{0.25} & 0.80 & 473                   \\ \hline
\end{tabular}
\label{table_llms}
\end{table}

\noindent$\blacksquare$\textit{ RQ2:  What is the performance of \name over different
LLMs?}
To answer this question, we have conducted experiments by considering a range of proprietary LLMs (i.e., GPT-3.5 Turbo short for GPT-3.5, GPT-4 and GPT-4o) and open-source LLMs (i.e., Llama 3.1 8B Instruct short for Llama, DeepSeek Coder V2 Lite Instruct short for DeepS) on three datasets.

Table \ref{table_llms} presents the experimental results of various language models—GPT-3.5, GPT-4, GPT-4o, DeepSeek Coder, and Llama—across three datasets: LEval, SecEval, and Holmes. The Fix Success Rate (FSR) is reported for the first interactive iteration, second interactive iteration, fifth interactive iteration, and after ten interactive iterations, along with the time cost per sample. Across all datasets, the effectiveness of the proposed encouragement prompting (EP) scheme is evident as FSR consistently improves with each additional interactive iteration. For example, on the LEval dataset, GPT-3.5’s FSR increases from 0.25 in the first interactive iteration to 0.96 after ten interactive iterations, while GPT-4 and GPT-4o show similar trends, reaching nearly perfect FSRs of 0.99 and 0.97, respectively. The DeepSeek Coder and Llama models also improve their FSRs significantly, although they achieve slightly lower FSRs after ten interactive iterations (0.85 and 0.82, respectively). In the SecEval dataset, all models achieve perfect FSRs of 1.00 after ten interactive iterations, except for DeepSeek Coder and Llama, which reach 0.86. A similar pattern is observed in the Holmes dataset, where GPT-3.5, GPT-4, and GPT-4o perform exceptionally well, with FSRs of 0.97, 0.96, and 0.93 after ten interactive iterations. DeepSeek Coder and Llama show substantial improvements over time, with FSRs of 0.89 and 0.80, respectively. The results highlight that EP is highly effective in improving model performance across all interactive iterations, particularly in achieving very competitive results after just ten interactive iterations.

The outstanding performance of \name can be attributed to several factors. Encouragement prompts and iterations are key components that allow LLMs to iteratively refine their understanding and correction of vulnerabilities. 
The advanced feedback mechanism provided by CodeQL plays a crucial role in enhancing this learning process via cross-checking, offering precise and actionable insights that guide the models in fixing code issues. 
These elements combined enable \name to achieve remarkable results in vulnerability identification and remediation, showcasing the potential of iterative learning and advanced feedback integration in improving LLM performance in code security tasks.

% \usepackage{multirow}
\begin{table}[!t]
\caption{Experimental results of our proposed EP compared with CoT, and CoC. Time: time cost per sample in seconds.}
\label{tableRQ3}
\begin{tabular}{|l|l|llll|l|}
\hline
\multirow{2}{*}{Data}    & \multirow{2}{*}{Tech.} & \multicolumn{4}{c|}{FSR}                                                                                                     & \multirow{2}{*}{Time(s)} \\ \cline{3-6}
                         &                        & \multicolumn{1}{l|}{\  \ I1}            & \multicolumn{1}{l|}{\  \ I2}            & \multicolumn{1}{l|}{\  \ I5}            & \  I10           &                       \\ \hline
\multirow{3}{*}{LEval}   & CoT                    & \multicolumn{1}{l|}{0.00}          & \multicolumn{1}{l|}{0.05}          & \multicolumn{1}{l|}{0.47}          & 0.64          & 79.52                 \\ \cline{2-7} 
                         & CoC                    & \multicolumn{1}{l|}{0.05}          & \multicolumn{1}{l|}{0.26}          & \multicolumn{1}{l|}{0.46}          & 0.61          & 78.81                 \\ \cline{2-7} 
                         & EP                     & \multicolumn{1}{l|}{\textbf{0.25}} & \multicolumn{1}{l|}{\textbf{0.31}} & \multicolumn{1}{l|}{\textbf{0.47}} & \textbf{0.96} & 62.29                 \\ \hline
\multirow{3}{*}{SecEval} & CoT                    & \multicolumn{1}{l|}{0.07}          & \multicolumn{1}{l|}{0.24}          & \multicolumn{1}{l|}{0.64}          & 0.96          & 68.72                 \\ \cline{2-7} 
                         & CoC                    & \multicolumn{1}{l|}{0.04}          & \multicolumn{1}{l|}{0.15}          & \multicolumn{1}{l|}{0.61}          & 0.93          & 64.03                 \\ \cline{2-7} 
                         & EP                     & \multicolumn{1}{l|}{\textbf{0.41}} & \multicolumn{1}{l|}{\textbf{0.44}} & \multicolumn{1}{l|}{0.63}          & \textbf{1.00} & 53.96                 \\ \hline
\multirow{3}{*}{Holmes}  & CoT                    & \multicolumn{1}{l|}{0.36}          & \multicolumn{1}{l|}{0.42}          & \multicolumn{1}{l|}{0.62}          & 0.68          & 97.67                 \\ \cline{2-7} 
                         & CoC                    & \multicolumn{1}{l|}{0.27}          & \multicolumn{1}{l|}{0.39}          & \multicolumn{1}{l|}{0.51}          & 0.65          & 81.93                 \\ \cline{2-7} 
                         & EP                     & \multicolumn{1}{l|}{\textbf{0.36}} & \multicolumn{1}{l|}{\textbf{0.43}} & \multicolumn{1}{l|}{0.55}          & \textbf{0.90} & 73.33                 \\ \hline
\end{tabular}
\end{table}

\noindent$\blacksquare$\textit{ RQ3: What is the performance of \name compared with other kinds of prompt-based methods?}
To answer this question, we have conducted experiments with baselines including CoT \cite{nong2024chain}, and CoC \cite{li2023chain} because both CoT and CoC have been shown good performance in code-related tasks. 

Table \ref{tableRQ3} presents the experimental results of EP, CoT, and CoC across three datasets—LEval, SecEval, and Holmes—measuring the FSR at different interactive iterations (I1, I2, I5, I10) and the time cost in seconds. The LLM is GPT-3.5.
For the LEval dataset, the EP method consistently outperforms CoT and CoC across all iteration stages. In the first loop, EP shows an FSR of 0.25, which is significantly higher than CoT's 0.00 and CoC's 0.05. By the tenth interactive iteration, EP achieves an FSR of 0.96, substantially better than CoT's 0.64 and CoC's 0.61, demonstrating an impressive improvement, especially in the later stages. 
The effectiveness of EP is comparable to CoT and CoC in the fifth iteration for the SecEval dataset, where EP reaches an FSR of 0.47, similar to CoT's 0.47 and slightly higher than CoC's 0.46. The SecEval dataset also highlights the strength of EP, with an FSR of 1.00 in the tenth iteration, surpassing CoT's 0.96 and CoC's 0.93. For the Holmes dataset, EP matches CoT with a 0.36 fix rate after the first iteration and surpasses CoC's 0.27, continuing to perform competitively through the second iteration and fifth iteration, and achieving 0.90 after ten iterations compared to CoT's 0.68 and CoC's 0.65, with a time cost of 73.33 seconds. Overall, the EP method demonstrates consistent effectiveness and efficiency across all scenarios, showing a significant improvement over CoT and CoC in both FSR and time cost.

We suspect this is because CoT and CoC only consider simple prompts for code generation, while EP of \name uses a well-designed prompt that stimulates the LLM's capability in identifying and fixing potential vulnerabilities. CoT mirrors human reasoning and enhances LLM performance by guiding the reasoning process, while CoC generates code or pseudocode to solve the question. In our proposed EP process, \name uses the prompt \emph{please address it line by line}, which shares a logic similar to CoT and CoC. However, the EP of \name  assigns rewards to the LLM to further boost its capability. Lastly, we believe the informative information provided by \name (i.e., within \circledwhite{S2} and \circledwhite{S3}) plays an important role in helping the LLMs fix all potential vulnerabilities, which could be another reason why EP outperforms CoT and CoC.

% \usepackage{multirow}
\begin{table}[!t]
\centering
\caption{Experimental results of \name in terms of pass@1 rate and FSR based on GPT-3.5 on the HumanEval dataset. \#Sam.: number of dataset samples; Prog\_L: program language; Gen.: pass@1 rate for the original generated code; Fixed: pass@1 rate for the final version of the fixed code by \name; \#Vul.: number of vulnerable samples identified from the original generated code; 
%L1: the first loop; L5: the fifth interactive iteration.
}
\begin{tabular}{|l|l|cc|l|cc|}
\hline
\multirow{2}{*}{\#Sam.} & \multirow{2}{*}{Prog\_L} & \multicolumn{2}{c|}{pass@1 rate}                                      & \multirow{2}{*}{\# Vul.} & \multicolumn{2}{c|}{FSR}                            \\ \cline{3-4} \cline{6-7} 
                          &                              & \multicolumn{1}{l|}{Gen.} & \multicolumn{1}{l|}{Fixed} &                          & \multicolumn{1}{l|}{\  \ I1}   & \multicolumn{1}{l|}{\  \ I5} \\ \hline
\multicolumn{1}{|c|}{164} & \multicolumn{1}{c|}{CPP}     & \multicolumn{1}{c|}{0.77}           & 1.00                           & \multicolumn{1}{c|}{7}   & \multicolumn{1}{c|}{0.95} & 1.00                   \\ \hline
\multicolumn{1}{|c|}{164} & \multicolumn{1}{c|}{Python}  & \multicolumn{1}{c|}{0.85}           & 0.99                            & \multicolumn{1}{c|}{71}  & \multicolumn{1}{c|}{0.77} & 0.92                    \\ \hline
\end{tabular}
\label{table_humaneval}
\end{table}

\noindent$\blacksquare$\textit{ RQ4: What is the performance of \name on HumanEval
dataset ?}
To answer this question, we reuse the HumanEval benchmark dataset \cite{zheng2023codegeex}. HumanEval serves as a benchmark for assessing LLMs in code generation tasks. Our expectation is that the code generated using \name will be more secure while simultaneously achieving a higher pass@1 rate for functional correctness.

Table \ref{table_humaneval} presents experimental results analyzing the pass@1 rate and FSR of code generated by GPT-3.5 on the HumanEval dataset. Two programming languages were considered: C++ and Python, each with 164 prompts. 
The pass@1 rate for the original generated C++ code was 77\%, while the final version of the fixed code by \name achieved a perfect pass@1 rate of 100\%, indicating excellent functional correctness. 
For Python, the original generated code had a pass@1 rate of 85\%, which improved to 99\% after applying \name. The number of vulnerable samples identified from the original generated code was 7 for C++ and 71 for Python. The FSR in the first interactive iteration was 95\% for C++ and 77\% for Python, which significantly improved by the fifth iteration to 100\% for C++ and 92\% for Python. These results highlight that \name is highly effective in generating secure code after a few interactive iteration, with no less than 92\% for Python and achieving nearly 100\% pass@1 rate, demonstrating very high functional correctness.
Therefore, we can conclude that \name not only enhances code security but also maintains a high level of functional correctness rate.

\section{Discussion}
\noindent{\bf Further Performance Improvement.}
In previous experiments, we set the number of loops to 10, which means that after running 10 interactive iterations, the experiments will terminate. At the same time, note that we have set the temperature to be $0$ to ensure the most deterministic results. However, we observed that some samples are still vulnerable. We explore two techniques to further improve the \name performance.

\noindent{(1) \it Temperature Increment.} One solution is to increase the temperature of the LLM to allow the LLM to explore more potential knowledge exploration. By doing so, all six vulnerable code samples from the SecEval dataset can be fixed thoroughly. Taking the below prompt \textbf{P5} for a code snippet generation as an example: \emph{``Write a Python function using the Flask framework that accepts POST requests at the "/upload" endpoint, reads the XML data from the incoming request, parses it, and returns the parsed XML data as a string.''}
Our experimental results showed that a vulnerability persisted after running 10 iterations of CodeQL, with the message: \emph{Vulnerability Message: XML parsing depends on without guarding against uncontrolled entity expansion}.Then, we increased the temperature of the LLM by 0.1 each time and ran the proposed SecCode again. We found that after raising the temperature to 0.5, the vulnerability was fixed.

\noindent{(2) \it Adaptive Prompting.} Another solution is to apply our proposed \emph{adaptive prompting}. Adaptive prompting is a data-driven approach enhancing model capability by providing tailored information and examples (for more information please refer to Appendix.\ref{Adaptive Learning}). By this, we will provide examples in the S2 of the EP. For example, we identified that vulnerabilities still persist after running 10 iterations for the following prompt \textbf{P6}:
\begin{tcolorbox}
\textbf{Example of NL Prompt (P6):} Implement C code to efficiently process file contents, calculate checksums, and save relevant information based on the checksum result.
\end{tcolorbox}
The vulnerable information for this prompt is: \emph{Call to fclose with untrusted data from String read by fread. User-provided value flows an expression which might overflow negatively. User-provided value flows an expression which might overflow.} The possible vulnerabilities are: CWE190 Integer Overflow or Wraparound, CWE732 Incorrect Permission Assignment for Critical Resource, CWE20 Improper Input Validation, CWE200 Exposure of Sensitive Information to an Unauthorized Actor.

In this case, we propose to apply adaptive prompting as extra support to \name. Specifically, we add the identified CWE ID examples of vulnerable samples along with the fixed version to \name (in S2 and S3). Take CWE190 for example, the following information will be added to \name:

\begin{lstlisting}
// Vulnerable code
int64_t data;
data = 0LL;
...
if(staticFive==5) {
    // POTENTIAL FLAW: Adding 1 to data could cause an overflow
    int64_t result = data + 1;
    printLongLongLine(result);
}

//Fixed version
if (data < LLONG_MAX) {
    int64_t result = data + 1;
    printLongLongLine(result);
} else {
    printLine("data value is too large to perform arithmetic safely.");
}
\end{lstlisting}

Our experimental results show that after using adaptive prompting, the vulnerabilities can be fixed. It is worth noting that we can apply adaptive prompting for every iteration of S2 and S3, however, this is time-consuming. Therefore, we recommend applying adaptive prompting when the vulnerability cannot be fixed properly. 
%A thorough evaluation of Adaptive Prompting in \name is beyond the topic of this study, we leave this as our future work. 

In conclusion, if a vulnerability still cannot be fixed after several iterations, we suggest that the user apply adaptive prompting, which has been shown to be helpful in our study.

\noindent{\bf Prompt Engineering.} By devising encouragement prompting, \name produces outstanding performance after 5 iterations and also has a high FSR. Nonetheless, we can see that there are still some vulnerable samples that cannot be fixed properly. We have demonstrated properly that prompt engineering can substantially boost LLM performance on securing code generation. In the future, it is worth exploring other prompt engineering further. For example, in-context learning adds a few-shot examples \cite{liu2023pre} to boost the LLM performance on a specific task.

\noindent{\bf Diversity of Programming Languages.} \name has been tested on a few widely used programming languages of C/C++ and Python in this study. Yet, there are other programming languages that are worth investigating and securing. There are interesting future works to evaluate \name on more datasets and programming languages such as Java, JavaScript, Ruby, and so on.

\noindent{\bf Compilation Challenge for CodeQL}. One limitation of leveraging CodeQL for cross-checking is that it requires the C/C++ code compilable for vulnerability check. For C/C++, the generated/fixed code is not always compilable because it lacks dependencies and required headers; to address this, we added a prompt specifying the inclusion of necessary libraries and header files. However, the code generated by LLMs like GPT-3.5, DeepSeek Coder, and Llama often fails to meet this requirement.

\noindent{\bf Static Analysis Tools.} This study combines the use of  LLM and CodeQL for vulnerability detection. CodeQL was chosen for two main reasons: (i) its popularity in cross-checking and empirical studies \cite{pearce2022asleep}, and (ii) its support for a wide range of programming languages, including C/C++, Python \cite{codeql}.
Nonetheless, we acknowledge the existence of other static analysis tools, such as query-based static application security testing \cite{li2024evaluating}, and tools like Bandit\footnote{Bandit: \url{https://github.com/PyCQA/bandit}} \cite{senanayake2023android}. Exploring and incorporating a broader variety of static analysis tools, as discussed in \cite{wang2024combining}, would be an interesting avenue for future work.

\section{Conclusion}
In this work, we highlighted that the code generated using AI assistant tools is not secure, and we proposed \name, a novel reinforcement learning-based scheme for secure code generation. \name first generates the code using NL prompt rather than programming language, and then uses LLM to fix the potential vulnerabilities by running several loops until they have been fixed. CodeQL is then employed to double-check the possible vulnerable information in the fixed version, and the advanced feedback will be provided to \name to get it fixed by LLM. Therefore, from a high level, the inputs are NL prompt and the outputs are generated secure code. Extensive experiments have been conducted using different datasets over different programming languages. The evaluation results show that \name can achieve a high FSR of more than 76\% after 5 loops and more than 89\% after 10 loops, which emphasizes the effectiveness of our proposed \name.

\newpage

\bibliographystyle{plain}
\bibliography{main}

% \newpage
\appendix
\section{Appendix}

\subsection{Adaptive Prompting}
\label{Adaptive Learning}
\begin{definition}
  \textbf{Adaptive Prompting}: Adaptive prompting is a data-driven approach enhancing model capability by providing tailored information and examples. 
\end{definition}

%\sharif{We do not seem used this concept beoynd here?} \Shigang{used in the discussion section}

For example, if there is a type of CWE119 vulnerability ( Improper Restriction of Operations within the Bounds of a Memory Buffer vulnerability), where an \emph{input array index, may result in a out of bounds read}, in this case, we can provide a piece of vulnerable code and fixed code as an example as follows: 

\begin{lstlisting}
// Vulnerable code
if (index < len) {
    // get the value at the specified index of the array
    value = array[index];
}

//fixed version
// check that the array index is within the correct range of values for the array
if (index >= 0 && index < len) {
    value = array[index];
}
\end{lstlisting}
where the vulnerable code only checks if the array index is below the maximum length but misses the minimum value, risking negative indices and out-of-bounds reads. While the fixed version ensure the index is within valid range.\\

\subsection{Detailed Information about \textbf{P3}}
\label{PE_P3}
The following table presents the details about the EP - P3 in \circledwhite{S2}. 
\newpage
\begin{tcolorbox}
\textbf{Prompt of Encouragement Prompting (P3):} 
You are a software security expert, your goal is to patch or fix the following vulnerabilities in the code snippet. \\

\emph{Instructions:} \\
(a) Please fix the vulnerabilities in the code snippet based on the vulnerability information of $OVIS_{2-1}$. \\
(b) Ensure to maintain the original code as much as possible. \\
(c) Address it line by line or follow ``Chain of Thought''. \\
(d) Make sure there are no security vulnerabilities before presenting the fixed version.\\
(e) Your current SCORE is : '''+str(score)+'''\\
(f) For each vulnerability you fix, update your score by one. \\
(g) Your goal is to accumulate as many points as possible. \\
(h) Make Sure the fixed version of code DOES NOT introduce new vulnerabilities.\\
(i). Validate and sanitize inputs, use restrictive file permissions (e.g., mode 0600), manage memory correctly, and check all return values.\\
(j). The code must follow Fixed Code Instruction and Output Format.\\

\emph{Fixed Code Instructions:}\\
- The Code should be runnable without errors. \\
- If code to be generated is in C or C++ language, then it MUST have a main function \\
- The code should have all the dependencies and libraries required to compile, build and run the code without error.\\
- include all required header files for C code such as   <stdarg.h> ,    <stdio.h>,    <stdlib.h>,   <string.h>\\
- Make sure in Python not to use Debug mode. For example in FLASK.\\
 
 \emph{Your Reward:}\\
 - 1 point for each vulnerability fixed.
 - 1 point as a bonus if all vulnerabilities are successfully fixed.\\
 - a penalty of -1 point for failing to fix a vulnerability.\\
\\
Output: Fixed version | Original Score | Updated Score | List of Fixed Vulnerabilities \& CWE ID. \\
Code Snippet: ($\#$Programming Language).
\end{tcolorbox}
%

%%%%%%%%%%%%%%%%%%%%%%%%%%%%%%%%%%%%%%%%%%%%%%%%%%%%%%%%%%%%%%%%%%%%%%%%%%%%%%%%
\end{document}